\documentstyle[aps,prl,twocolumn,epsf]{revtex}

\begin{document}

\twocolumn[
\hsize\textwidth\columnwidth\hsize\csname@twocolumnfalse\endcsname
\draft

\title{Normal-state magnetotransport in 
La$_{1.905}$Ba$_{0.095}$CuO$_{4}$ single crystals}
\author{Yasushi Abe$^{1,2}$, Yoichi Ando$^{1,2}$, J. Takeya$^{1}$, 
H.Tanabe$^{3}$, T. Watauchi$^{3}$, I. Tanaka$^{3}$, and H. Kojima$^{3}$}
\address{$^{\rm 1}$ Central Research Institute of Electric Power
Industry, Komae, Tokyo 201-8511, Japan}
\address{$^{\rm 2}$ Department of Physics, Science University of Tokyo, 
Shinjuku-ku, Tokyo 162-8601, Japan}
\address{$^{\rm 3}$ Institute of Inorganic Synthesis,
Yamanashi University, Kofu, Yamanashi 400-8511, Japan}
\date{Received LBCO-3f.tex}
\maketitle

\begin{abstract}
The normal-state magnetotransport properties of 
La$_{2-x}$Ba$_{x}$CuO$_{4}$ single crystals
with $x$=0.095 are measured; at this composition, 
a structural transition to a low-temperature 
tetragonal (LTT) phase occurs {\it without} suppression of 
superconductivity.
None of the measured properties (in-plane and out-of-plane resistivity, 
magnetoresistance, and Hall coefficient) shows any sudden change
at the LTT phase transition, indicating that 
the occurrence of the LTT phase does not necessarily cause an 
immediate change in the electronic state such as the charge-stripe 
stabilization.
\end{abstract}

\pacs{PACS numbers: 74.25.Fy, 74.62.Bf, 74.25.-q, 74.72.Dn}

]

La$_{2-x}$Ba$_{x}$CuO$_{4}$ (LBCO) has been generally considered as a 
rather peculiar high-$T_c$ cuprate, not only because it is the first 
high-$T_c$ cuprate discovered by Bednorz and M\"uller \cite{Bednorz},
but also because $T_c$ of this compound is drastically suppressed in the 
composition rage near $x$=1/8 \cite{Moodenbaugh}, 
known as the \lq \lq 1/8 anomaly".
Soon after the 1/8 anomaly was recognized, 
it was found that LBCO system shows 
a structural phase transition from a low-temperature orthorhombic (LTO) 
phase to a low-temperature tetragonal (LTT) phase
in a rather wide range of $x$ around 1/8 \cite{Axe,Suzuki}.
On the other hand, La$_{2-x}$Sr$_{x}$CuO$_{4}$ (LSCO) system, 
which has the same crystal structure as LBCO, does not show a
clear suppression of $T_c$ near 1/8;
since there is no structural transition to the LTT phase in LSCO
\cite{Takagi,Yamada}, 
it is generally believed that the occurrence of the LTT phase is 
responsible for the suppression of $T_c$ in LBCO.

There have been many experiments which tried to investigate the
fundamental mechanism of the 1/8 anomaly.
For example, Yoshida {\it et al.} studied the effect of partial 
substitution of Ba$^{2+}$ ion in LBCO with smaller
divalent cations and found that such 
replacement of Ba$^{2+}$ leads to
a suppression of the LTT structural transition and 
simultaneously to a recovery of the superconductivity \cite{Yoshida}.
This result suggests that the LTT transition temperature and the 
strength of the $T_c$ suppression are closely tied to each other.
Thus, Yoshida {\it et al.} concluded that the \lq\lq 1/8 anomaly" is 
caused by a 
Peierls-type mechanism with cooperative electronic and lattice instabilities.
However, there are evidences which suggests that the 
occurrence of the LTT phase alone does not necessarily mean a
destruction of superconductivity.
Behaviors of La$_{2-x-y}$Nd$_{y}$Sr$_{x}$CuO$_{4}$ 
(Nd-doped LSCO) system is one such example \cite{Nakamura}.
In this system, while there is a clear structural phase 
transition to the LTT phase at 71 K and the superconductivity is
almost completely destroyed for $x$=0.12, 
there remains a bulk superconductivity (with $T_c$=16 K)
for $x$=0.20 even though the LTT phase transition temperature
$T_{LTT}$ is higher (79 K) than that for $x$=0.12.
Since high-quality single crystals are available for Nd-doped LSCO,
the in-plane resistivity $\rho_{ab}$ and the out-of-plane 
resistivity $\rho_c$ have been studied in this system \cite{Nakamura}.
For $x$=0.12, both $\rho_{ab}$ and $\rho_c$ show a clear jump
at $T_{LTT}$, suggesting that the electronic state is changed
upon the structural phase transition.
It was found that $\rho_c$ shows a jump at $T_{LTT}$ even for $x$=0.20, 
indicating that the change in the electronic state persists to the $x$ 
value where the suppression of superconductivity is weak.

The known properties of LBCO is quite similar to that of Nd-doped LSCO;
superconductivity is almost completely destroyed at $x$=1/8,
bulk superconductivity remains for $x$$\neq$1/8,
and the structural phase transition to the LTT phase occurs around
60 K which is almost independent of $x$. 
However, because of the difficulty in growing single crystals of LBCO,
the anisotropic resistivity and the magnetotransport properties 
have not been well studied in LBCO with $x$ near 1/8 and thus
the electronic states near $x$=1/8 is not well understood.

One of the composition of particular interest in LBCO is $x$=0.09;
around this composition, the structural phase transition takes place
but $T_c$ is not suppressed ($T_c$$\simeq$30 K).
In other words, the superconductivity for $x$=0.09 does not seem to be 
affected by the occurrence of the LTT phase.
It is thus interesting to study whether the electronic system 
shows any change at the LTT phase transition for $x$=0.09,
where the LTT phase does not affect superconductivity at all.
This may clarify the importance (or unimportance) of the occurrence
of the LTT phase to the electronic structure.

With the improvement in the crystal growth technique, 
high-quality single crystals of LBCO with $x$ near 1/8 have recently
become available \cite{Tanabe,Khan}.
In this paper, we report our detailed measurement of the anisotropic
normal-state resistivity ($\rho_{ab}$ and $\rho_c$), 
in-plane magnetoresistance (MR), and the Hall coefficient $R_H$,
of LBCO single crystals with $x$=0.095.
As discussed above, this is the particular composition where 
$T_c$ is not suppressed despite the presence of the LTT phase.
In fact, our $x$=0.095 crystals showed mid-point $T_c$ of 31 K, 
a very high value for LBCO.
It was found that none of the measured transport properties shows
any drastic change at the LTT phase transition, which strongly
support the picture that the occurrence of the LTT phase does not
necessarily change the electronic system.

The question whether the occurrence of the LTT phase alone
can be responsible for the change in the electronic state is
particularly intriguing in the light of the recently reported
\lq \lq stripe order" in the Nd-doped LSCO with $x$=0.12;
using neutron diffraction techniques, 
Tranquada {\it et al.} observed elastic magnetic superlattice peaks of 
the type (1/2$\pm \epsilon$,1/2,0)
and charge-order peaks at (2$\pm$2$\epsilon$,0,0), where $\epsilon$=0.118 
at low temperatures \cite{Tranquada,Tranquada2}.
Such an observation strongly suggests a presence of a one-dimensional
charge order (\lq \lq stripes") which intervene in the antiferromagnetic
spin order. 
Tranquada {\it et al.} proposed that 
the modulated antiferromagnetic order is 
pinned and stabilized in the LTT phase but not in the LTO phase,
which is the reason why such static structure is not observed 
in pure LSCO.
Following this picture, it can be inferred that the fundamental
origin of the change in the electronic state in Nd-doped LSCO is
the occurrence of the stripe phase and not the occurrence of the
LTT phase itself.
If so, it may be that the stripe order is {\it not} stabilized by the
LTT phase transition in LBCO at $x$=0.095, 
which can be the reason for the coexistence
of a \lq \lq high" $T_c$ of 31 K with the LTT phase. 

The single crystals of La$_{1.905}$Ba$_{0.095}$CuO$_{4}$ are grown 
using a traveling-solvent floating-zone (TSFZ) technique.
Details of the crystals growth of LBCO are described 
elsewhere \cite{Tanabe}.
After the crystallographic axis are determined, we cut the crystals 
to sufficiently small dimensions, typically 2 $\times$ 0.4 $\times$
0.1 mm$^3$, to ensure homogeneous Ba concentration in
the crystal.
The crystals are annealed in flowing-oxygen atmosphere
at 650$^{\circ}$C for 24 hours to remove oxygen deficiencies.  
The actual Ba concentrations in the crystals are determined by the 
inductively-coupled plasma spectrometry (ICP) technique.
A standard six-terminal method is used for the simultaneous
$\rho_{ab}$ and $R_H$ measurement.
Both the MR and $R_H$ data are taken in the sweeping 
magnetic field at fixed temperatures with an ac technique.
The temperature is very carefully controlled and stabilized using 
both a capacitance sensor and a Cernox resistance sensor
to avoid systematic temperature deviations with magnetic fields.
The stability of the temperature during the MR and $R_H$
measurements is within 10 mK.

Figure 1 shows the temperature dependence of $\rho_{ab}$ and 
$\rho_{c}$.  These data are measured in two different samples
cut from the same rod.
In both samples, the onset $T_c$ is 33 K and the resistivity becomes 
zero at 29 K.
$\rho_{ab}$ is linear in $T$ down to 150 K and
shows an upward deviation from the $T$-linear behavior 
at lower temperatures. 
A slight upturn in $\rho_{ab}$ is observed below 45 K, which is
consistent with the data on polycrystalline samples around
this composition \cite{Uchida}.
The extrapolated residual resistivity is negligibly small, which is
similar to the behavior of high-quality LSCO crystals \cite{Kimura}. 
In the case of Nd-doped LSCO, clear jumps in both $\rho_{ab}$ and
$\rho_c$ have been observed at $T_{LTT}$ in underdoped samples 
\cite{Nakamura};
however, there is no clear jump neither in $\rho_{ab}$ nor in 
$\rho_c$ in LBCO as shown in Fig. 1.
Note that the structural phase transition to the LTT phase takes place at
about 60 K for this $x$ value in LBCO \cite{Phillips}.
Therefore, contrary to the Nd-doped LSCO system, 
the resistivity data suggest that there is no sudden change 
in the electronic system in LBCO with $x$=0.095 at $T_{LTT}$.
If we look at the temperature dependence of 
$d\rho_{ab}/dT$ (Fig. 1 inset, upper curve),
there is a kink near $T_{LTT}$ ($\simeq$60 K), 
which may suggest that the
scattering of electrons gradually increases in the LTT phase.
On the other hand, $d\rho_c/dT$ (Fig. 1 inset, lower curve) 
does not show any change at $T_{LTT}$, although there is a kink
at lower temprature, about 52 K.
It is intriguing that $d\rho_{ab}/dT$ and $d\rho_c/dT$ show kinks
at different temperatures. 

\vspace{-0.7cm}
\begin{figure}[htbp]
\begin{center}
 \epsfxsize=75mm
 $$\epsffile{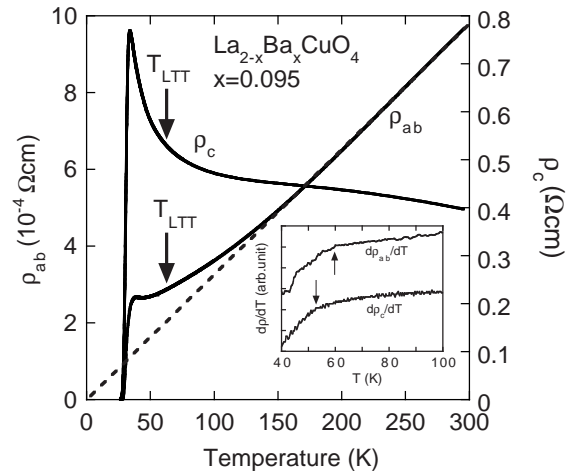}$$
\end{center}
\caption{$T$ dependence of $\rho_{ab}$ (left-hand-side axis) 
and $\rho_c$ (right-hand-side axis).
$T_{LTT}$ is indicated by arrows.
Inset: Plot of $d\rho_{ab}/dT$ and $d\rho_c/dT$ vs $T$.
Arrows mark the kinks.}
\label{fig1}
\end{figure}

Figure 2 shows the temperature dependence of the in-plane $R_H$.
Here, the magnetic field is applied along the $c$ axis and the
current is along the $ab$ plane.
For comparison, $R_{\rm H}$ data of LSCO ($x$=0.1) polycrystalline sample 
\cite{Hwang} are also shown by a dashed line.
$R_H$ of LBCO shows a peak at about 50 K, which is nearly the same
temperature where $\rho_{ab}$ starts to show an upturn. 
There is no appreciable change in $R_H$ at $T_{LTT}$ (=60 K).
The behavior and the absolute value of the $R_H$ of our LBCO crystal 
are quite similar to that of LSCO ($x$=0.1) system, 
which does not show an LTT phase transition. 

One popular way of analyzing the 
normal-state transport properties of cuprates is to 
consider two scattering times, $\tau _{tr}$ and $\tau _H$ \cite{Iye}.
$\tau _{tr}(T)$ and $\tau _{H}(T)$ are determined by the 
temperature dependence of $\rho_{\rm ab}$ and the cotangent of the 
Hall angle $\theta _{\rm H}$, respectively \cite{Chien}.
Figure 3 shows $\cot \theta _{\rm H}$ (=$\rho_{xx}$/$\rho_{xy}$)
at 10 T plotted against $T^2$.
Since the Hall angle is proportional to the inverse of $\tau _{\rm H}$,
it is clear from Fig. 3 that $\tau _{\rm H}^{-1}$ obeys a $T^2$ law
very well across $T_{LTT}$ down to 45 K.
(The inset to Fig. 3 is a modified plot of the main panel to
show directly the temperature region where the $T^2$ law holds.)

\vspace{-0.7cm}
\begin{figure}[htbp]
\begin{center}
 \epsfxsize=75mm
 $$\epsffile{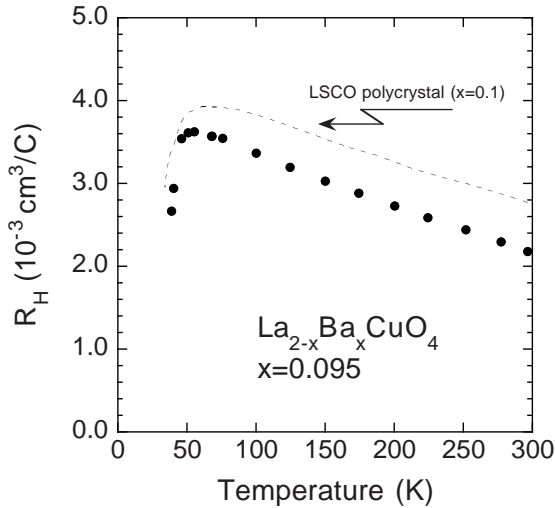}$$
\end{center}
\caption{$T$ dependence of the Hall coefficient 
$R_{\rm H}$(T) (solid circles).
The dashed line is the $R_H(T)$ data of LSCO ($x$=0.1) polycrystalline
sample from Hwang {\it et al.} [16].}
\label{fig2}
\end{figure}

\vspace{-0.7cm}
\begin{figure}[htbp]
\begin{center}
 \epsfxsize=75mm
 $$\epsffile{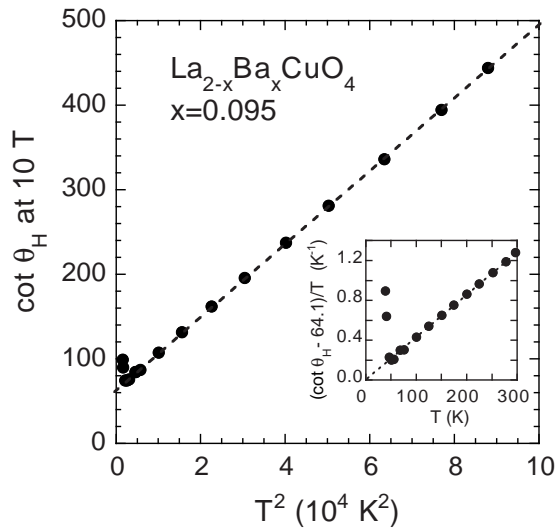}$$
\end{center}
\caption{$T^{\rm 2}$ plot of $\cot \theta_{H}$ at 10 T. 
The dashed line is a fit to the data with 
$\cot \theta _{\rm H}=a+bT^2$ ($a$=32.1 and $b$=21.3). 
Inset: A modified plot of the main panel to
show that the $T^2$ law holds down to 45 K.}
\label{fig3}
\end{figure}

Figure 4 shows the result of the MR measurements of the LBCO crystal.
We measured both the transverse MR ($I$ is within the $ab$ plane 
and $H$ is parallel to the $c$ axis) and the longitudinal MR 
($I$ and $H$ are within the $ab$ plane and $H$ is parallel to $I$).
The transverse MR consists of orbital and spin contributions,
while the longitudinal MR comes only from the spin contribution.
By comparing the two MRs, we can see that the spin contribution to
the transverse MR is not large (about 30\%).
Although the longitudinal MR shows a smooth increase down to 40 K, the 
transverse MR shows a rather steep increase below 60 K, resulting
in more than an order-of-magnitude difference between the two MRs 
at 40 K.

\vspace{-0.7cm}
\begin{figure}[htbp]
\begin{center}
 \epsfxsize=75mm
 $$\epsffile{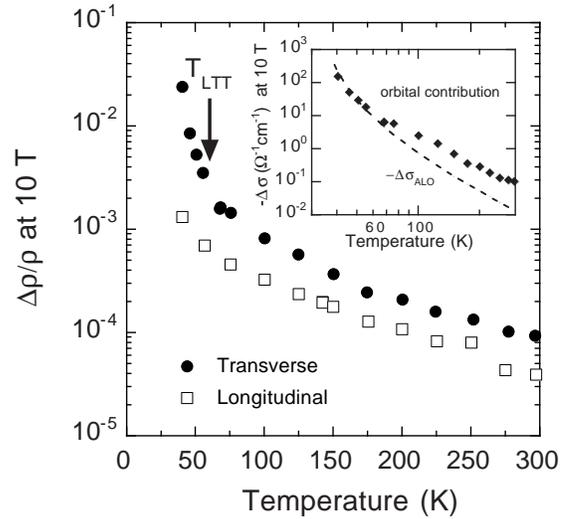}$$
\end{center}
\caption{$T$ dependence of the transverse MR (sold circles) 
and the longitudinal MR (open squares) at 1 T.
The arrow shows $T_{LTT}$.
Inset: Orbital part of MR and the estimated AL orbital 
fluctuation conductivity (dashed line).}
\label{fig4}
\end{figure}

We tried to analyze whether this steep enhancement in the transverse MR
can be understood by the superconducting fluctuation conductivity,
whose contribution is large only for the transverse geometry.  
The fluctuation conductivity consists of Aslamasov-Larkin (AL) term and 
Maki-Thompson (MT) term; both terms comprises two contributions, 
the orbital contribution and the spin contribution \cite{Iye}.
Kimura {\it et al.} have analyzed the MR in underdoped 
LSCO and concluded that the
MT term is absent \cite{Kimura}, which is actually expected for a 
$d$-wave superconductor \cite{Yip}.
Thus we tried to estimate the fluctuation conductivity by considering
only the AL term.
The dashed line in the inset to Fig. 4 is the estimated AL orbital
contribution, where the parameters are 
$\xi_{\rm ab}(0)$=30 {\AA} and $\xi_{\rm c}(0)$=1 {\AA}.
(We just assumed these values as typical values.)
The orbital part of MR, which is obtained by subtracting the longitudinal
MR from the transverse MR, is also plotted in Fig. 4.
Clearly, the increase of the transverse MR below 60 K can 
be accounted for by the superconducting fluctuations;
therefore, it is not likely that the steep increase in the
transverse MR is related to the occurrence of the LTT phase.

It has been proposed that the orbital MR in high-$T_c$ cuprates reflects
the variance of a local Hall angle around the Fermi surface and therefore
is proportional to the square of $\theta _{\rm H}$ \cite{Harris}, 
which is sometimes called \lq \lq modified Kohler's rule".
Figure 5 shows the temperature dependence of the orbital MR plotted 
together with $a\times (\cot \theta _{\rm H})^{\rm -2}$, where
$a$ is a fitting parameter.
[Note that $(\cot \theta_{\rm H})^{\rm -2}$$\simeq$$\theta_{\rm H}^2$
when $\theta_{\rm H}$ is small.]
The orbital MR does not scale so well to $(\cot \theta _{\rm H})^{\rm -2}$.
Particularly, the orbital MR shows weaker temperature dependence above 
$\sim$200 K compared to $(\cot \theta _{\rm H})^{\rm -2}$.
This might be an indication that the modified Kohler's rule is not
universally applicable to all high-$T_c$ cuprate.
It would be interesting to study the applicability of the 
modified Kohler's rule to the LBCO system in a wider 
carrier-concentration range.

\vspace{-0.7cm}
\begin{figure}[htbp]
\begin{center}
 \epsfxsize=75mm
 $$\epsffile{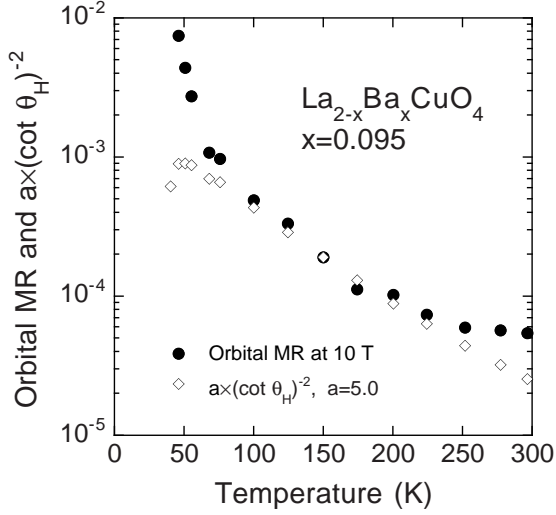}$$
\end{center}
\caption{Orbital MR and $a\times (\cot \theta_H)^{-2}$ vs $T$.}
\label{fig5}
\end{figure}

The above results indicate altogether that the electronic system
as inferred from $\tau _{\rm H}$ and $\tau_{tr}$ does not
show any sudden change at the LTT phase transition, which seems to be
different from the result of Nd-doped LSCO \cite{Nakamura}.
In particular, the fact that $\cot \theta _{\rm H}$ shows a 
good $T^2$ behavior down to 45 K (Fig. 3) 
suggests that $\tau _{\rm H}$ is not influenced by
the LTT phase.  On the other hand, $\tau_{tr}^{-1}$ seems to
grow gradually with lowering temperature in the LTT phase, 
which causes a faster increase in resistivity.
Based on these observations, we may conclude that
the coexistence of a \lq \lq high" $T_c$ of 31 K with the LTT phase
is possible in LBCO at $x$=0.095 because the LTT phase transition
does not immediately affect the electronic system.
This, however, does not rule out the possibility that the
electronic system is gradually changed in the LTT phase.
The localization behavior in $\rho_{ab}$ below 45 K might actually
be the result of some gradual change in the electronic state.

The authors would like to acknowledge Prof. S. Uchida for
valuable discussions and for showing us unpublished results
on polycrystalline LBCO.

%


\begin{references}
\bibitem{Bednorz}
J. G. Bednorz and K. A. M\"uller, Z Phys. B 64, 189 (1986). 
\bibitem{Moodenbaugh}
A. R. Moodenbaugh, Youwen Xu, and M. Suenaga, Ohys. Rev. B 38, 4596 (1988).
\bibitem{Axe}
J. D. Axe, A. H. Moudden, D. Hohlwein, D. E. Cox, K. M. Mohanty, A. R. 
Moodenbaugh and Y. Xu, Phys. Rev. Lett. 62, 2751 (1989).
\bibitem{Suzuki}
T. Suzuki and T. Fujita, Physica C 159, 111 (1989).
\bibitem{Takagi}
H. Takagi, T. Ido, S. Ishibashi, M. Uota, S. Uchida and Y. Tokura, Phys. 
Rev. B 40, 2254 (1989).
\bibitem{Yamada}
N. Yamada and M. Ido, Physica C 203, 240 (1992).
\bibitem{Yoshida}
K. Yoshida, F. Nakamura, Y. Tanaka, Y. Maeno and T. Fujita, Physca C 230, 
371 (1994).
\bibitem{Nakamura}
Y. Nakamura and S. Uchida, Phys. Rev. B 46, 5841 (1992).
\bibitem{Tanabe}
H. Tanabe, S. Watauchi, I. Tanaka, H. Kojima, {\it Advances in 
Superconductivity X},
Proceedings of the 10th International Symposium on Superconductivity Vol. 
1, 371 (1997).
\bibitem{Khan}
M.K.R. Khan, H. Tanabe, I. Tanaka, and H. Kojima,
Physica C 258, 315 (1996).
\bibitem{Tranquada}
J. M. Tranquada, B. J. Sternlieb, J. D. Axe, Y. Nakamura and S. Uchida, 
Nature 375, 561 (1995).
\bibitem{Tranquada2}
J. M. Tranquada, J. D. Axe, N. Ichikawa, Y. Nakamura, and S. Uchida, 
Phys. Rev. B 54, 7489 (1996).
\bibitem{Uchida}
S. Uchida (unpublished).
\bibitem{Kimura}
T. Kimura, S. Miyasaka, H. Takagi, K. Tamasaku, H. Eisaki,
S. Uchida, K. Kitazawa, M. Hiroi, M. Sera, and N. Kobayashi, Phys. Rev. B 
53, 8733 (1996).
\bibitem{Phillips}
J.C. Phillips and K.M. Rabe, Phys. Rev. B 44, 2863 (1991), and
refs. therein.
\bibitem{Hwang}
H.Y. Hwang, B. Batlogg, H. Takagi, H. L. Kao, J. Kwo, R. J. Cava, J. J. 
Krajewski and W. F. Peck, 
Phys. Rev. Lett. 72, 2636 (1994).
\bibitem{Iye}
For a review see Y. Iye, in {\it Physical Properties of High Temperature 
Superconductors III},
edited by D. M. Ginsberg (World Scientific, Singapore, 1992).
\bibitem{Chien}
T. R. Chien, Z. Z. Wang and N. P. Ong, Phys. Rev. Lett. 67, 2088 (1991).
\bibitem{Yip}
S. K. Yip, Phys. Rev. B 41, 2612 (1990).
\bibitem{Harris}
J. M. Harris, Y. F. Yan, P. Matl, N. P. Ong, P. W. Anderson, T. Kimura 
and K. Kitazawa, Phys. Rev. Lett. 75, 1391 (1995).

\end{references}
\end{document}